# The 1905 Relativity Paper and the "Light Quantum"

Galina Weinstein

Program for Science & Technology Education

Ben Gurion University, Israel

In 1905 the well-known physicist Max Planck was coeditor of the *Annalen der Physik*, and he accepted Einstein's paper on light quanta for publication, even though he disliked the idea of "light quanta". Einstein's relativity paper was received by the *Annalen der Physik* at the end of June 1905 and Planck was the first scientist to notice Einstein's relativity theory and to report favorably on it. In the 1905 relativity paper Einstein used a seemingly conventional notion, "light complex", and he did not invoke his novel quanta of light heuristic with respect to the principle of relativity. He chose the language "light complex" for which no clear definition could be given. But with hindsight, in 1905 Einstein made exactly the right choice not to mix concepts from his quantum paper with those from his relativity paper. He focused on the solution of his relativity problem, whose far-reaching perspectives Planck already sensed.

In his 1905 paper Einstein considers the system K. Very far from the origin of K, there is a source of electromagnetic waves. Let part of space containing the origin of coordinates be represented to a sufficient degree of approximation by plane waves:[1]

$X = X_0 \sin \Phi, Y = Y_0 \sin \Phi, Z = Z_0 \sin \Phi,$

$L = L_0 \sin \Phi, M = M_0 \sin \Phi, N = N_0 \sin \Phi.$

The phase is: $\Phi = \omega[t - (ax + by + cz)/c]$

$(X_0, Y_0, Z_0)$, $(L_0, M_0, N_0)$ are the components of the amplitude of the plane wave and a, b, c are the direction cosines and wave-normal **n**; $\omega$ is the frequency of the plane waves (really a sphere, but the radius is so big that it can be treated as plane waves). Einstein asks: What characterizes the waves when they are examined by an observer at the same point 0, but at rest in the system k?

Einstein applies the Lorentz transformation and transformation equations for electric and magnetic fields to the equations of the plane electromagnetic wave with respect to K. He obtains a new set of equations, from which he deduces new transformation equations for the frequency $\omega$ and direction cosines of the wave normal **n'**; Einstein obtains the Doppler principle and the equation that expresses the law of aberration.

---

[1] Einstein, Albert, "Zur Elektrodynamik bewegter Körper, *Annalen der Physik* 17, 1, 1905, pp. 891-921; pp. 910-911. (equations are written in original notation).

The Doppler principle for any velocity is the following: If an observer is moving with velocity v relative to an infinitely distant source of light of frequency ν, in such a way that the connecting line 'light source-observer' forms an angle φ with the velocity of the observer, which is referred to the system K relative to the source of light, the frequency ν' of the light perceived by the observer, is given by the equation":[2]

$$\nu' = \nu \frac{1 - \cos\varphi\, v/c}{\sqrt{1 - v^2/c^2}}$$

When φ = *0* the equation assumes the simple following form:

$$\nu' = \nu \sqrt{\frac{1 - v/c}{1 + v/c}}$$

Einstein then finds the amplitude of the waves as it appears in the system k; the amplitude of the electric or magnetic waves A or A', respectively, as it is measured in the system K or in the system k. He thus obtains:

$$A'^2 = A^2 \frac{(1 - \cos\varphi\, v/c)^2}{1 - v^2/c^2}$$

For φ = 0 this equation reduces to the simple form:

$$A'^2 = A^2 \frac{1 - v/c}{1 + v/c}$$

Einstein gives the equation for the square of amplitude, intensity $\propto A^2$.

Einstein then obtains the transformation of the energy of light rays. He could use the above result in the following way: if $A^2/8\pi$ is equal to the energy of light per unit volume $E = (E^2 + B^2)/8\pi$ measured in K, then according to the principle of relativity $A'^2/8\pi$ would be the energy of light per unit volume as measured in system k. Hence, we expect that the ratio, $A'^2/A^2$ would be the energy of a given light complex "measured in motion" and "measured at rest", if the volume of a light complex were the same measured in K and k. However, says Einstein, *this is not the case*.

Einstein thus instead considers a spherical surface of radius R moving with the velocity of light; a, b, and c are the direction-cosines of the normal to the light in the system K. Einstein is interested in the light energy enclosed by the light surface. No energy passes outside through the surface of the spherical light surface, because the surface and the light wave both travel with the velocity of light.

---

[2] Einstein, 1905, p. 911.

Einstein calculates the amount of energy enclosed by this surface as viewed from the system k, which will be the energy of the light complex relative to the system k.[3]

The spherical surface – viewed in the system k – is an ellipsoidal surface. If S is the volume of the sphere, and S' that of this ellipsoid, then he shows that:

$$\frac{S'}{S} = \frac{\sqrt{1 - v^2/c^2}}{1 - \cos\varphi\, v/c}$$

If we call the energy of the light enclosed by this surface E when it is measured in system K, and E' when measured in system k, we obtain the equation that relates between E and E':

$$\frac{E'}{E} = \frac{A'^2 S'/8\pi}{A^2 S/8\pi} = \frac{1 - \cos\varphi\, v/c}{\sqrt{1 - v^2/c^2}}$$

And when $\varphi = 0$ it simplifies to:[4]

$$\frac{E'}{E} = \sqrt{\frac{1 - v/c}{1 + v/c}}$$

Einstein writes, "It is noteworthy that the energy and the frequency of a light complex vary with the observer's state of motion according to the same law".[5]

Namely, $E'/E = \nu'/\nu$.

John Stachel says that this formula corresponds to that of the light quantum hypothesis, and in hindsight this supplies extra evidence for the later hypothesis. Einstein's aim is to show that the equation $E = h\nu$ that he uses in the quantum paper takes the same form in any inertial frame. $E = h\nu$ is transformed to $E' = h\nu'$ and thus the relativity postulate is not violated.[6]

Robert Rynasiewicz writes, "Einstein wraps up his derivation with what is clearly an allusion to the light quantum hypothesis". Rynasiewicz adds that "What he does not draw attention to there is the intimate relation of this result to the relative character of simultaneity".[7]

In his *Autobiographical notes* Einstein explained that while working simultaneously on the quantum problem and the nature of radiation, and on the electrodynamics of moving bodies, he "came to the conviction that only the discovery of a universal formal principle [the relativity principle] could lead us to assured results".[8]

---

[3] Einstein, 1905, p. 913.
[4] Einstein, 1905, pp. 913-914.
[5] Einstein, 1905, p. 914.
[6] Stachel, John, "Albert Einstein", *The New Dictionary of Scientific Biography*, Vol. 2, Gale 2008, pp. 363-373; p. 365.
[7] Rynasiewicz, Robert, "The optics and electrodynamics of 'On the Electrodynamics of Moving Bodies'", in Renn, Jürgen, (ed.) *Einstein's Annalen Papers. The Complete Collection 1901-1922*, Germany: Wiley-VCH Verlag GmbH& Co, 2005, pp. 38-57; p. 47.

Before submitting his 1905 special relativity paper, Einstein had submitted a paper on what came to be called his light quantum hypothesis – the only one of his 1905 papers he considered truly revolutionary: "On a Heuristic Viewpoint Concerning the Generation and Transformation of Light", sent to the *Annalen* on March 17$^{th}$, 1905, and received by the *Annalen* a day afterwards.[9] Einstein wrote Conrad Habicht in May 1905 about this paper, "It deals with the radiation and energy characteristics of light and is very revolutionary".[10]

This paper extended the range of application of Planck's 1900 quantum hypothesis. In order to explain his law of black body radiation, which had been well-verified empirically, Planck was forced to assume that oscillators interacting with the electromagnetic field could only emit and/or absorb energy in discrete units, which he called quanta of energy. The energy of these quanta was proportional to the frequency of the oscillator: $E = h\nu$. But Planck believed, in accord with Maxwell's theory, that the energy of the electromagnetic field itself could change continuously.

Einstein now showed that, if this formula were extended to the electromagnetic field energy itself, a number of phenomena involving interactions between matter and radiation, otherwise inexplicable classically, could now be simply explained with the help of these light quanta.

But, he was at work on his relativity paper too; so the question naturally arose, if the equation $E = h\nu$ holds in one inertial frame of reference, will it hold in all others. If not, then Einstein's relativity principle would be violated. Since h, the so-called quantum of action, is a universal constant, the question reduces to: Do the energy and frequency of a light quantum transform in the same way in passing from one inertial frame to another. And this is just what he demonstrates in his paper.

Hence, not wanting to introduce a discussion of his still-quite-speculative light quantum hypothesis into a paper which he regarded as simply an extension of well accepted classical ideas from mechanics to electromagnetism and optics, he confined his proof to the classical level.

---

[8] Einstein, Albert ,"Autobiographical notes" In Schilpp, Paul Arthur (ed.), *Albert Einstein: Philosopher-Scientist*, 1949, La Salle, IL: Open Court, pp. 1–95; pp. 48-49.
[9] Einstein Albert, "Über einen die Erzeugung und Verwandlung des Lichtes betreffenden heuristischen Gesichtspunkt", *Annalen der Physik* 17, 1905, pp. 132-148.
[10] Einstein to Habicht, 18 or 25 May, 1905, *The Collected Papers of Albert Einstein. Vol. 5: The Swiss Years: Correspondence, 1902–1914* (*CPAE*, Vol. 5), Klein, Martin J., Kox, A.J., and Schulmann, Robert (eds.), Princeton: Princeton University Press, 1993, Doc. 27.

Instead of "light quanta", in his proof he introduced the rather awkward term "light complex", a term that he soon dropped.[11]

In fact, Planck was the first scientist to notice Einstein's relativity paper. Einstein's paper on relativity, received by the *Annalen der Physik* at the end of June 1905 was already in print by 26 September. And as early of November 1905 Planck had reported favorably on it.[12]

However, Planck disliked the idea of light quanta. He was coeditor of the *Annalen*, and he accepted Einstein's "heuristic" paper on light quanta for publication, even though he objected to its basic concepts.

Einstein later wrote his friend Jacob Johann Laub about Planck,[13]

"He has, however, one fault: that he is clumsy in finding his way about in foreign trains of thought. It is therefore understandable when he makes quite faulty objections to my latest work on radiation" (light quantum).

In 1906 Planck's assistant, Max Laue, wrote Einstein on obtaining the preprints of the 1905 light quanta paper,

"When at the beginning of your last paper, you formulate your heuristic standpoint to the effect that radiant energy can be absorbed and emitted only in specific finite quanta, I have no objections to make; all of your applications also agree with this formulation. Now, this is not a characteristic of electromagnetic process in vacuum but rather of the emitting or absorbing matter, and hence radiation does not consist of light quanta as it says in §6 of your first paper; rather, it is only when it is exchanging energy with matter that it behaves as if it consisted of them."

Laue ended his letter to Einstein by saying: "By the way, I have never discussed your heuristic point of view with my boss. It is possible that there are differences of opinion between him and me on this question." [14]

But indeed the boss did agree with his assistant.

The boss wrote Einstein on July 6, 1907: [15]

"In any case, I do not believe that this difference in our opinion is of a fundamental nature.

---

[11] *The Collected Papers of Albert Einstein. Vol. 2: The Swiss Years: Writings, 1900–1909* (*CPAE*, Vol. 2), Stachel, John, Cassidy, David C., and Schulmann, Robert (eds.), Princeton: Princeton University Press, 1989, Doc. 23, note 31; note 39, p. 309.
[12] Hoffmann Banesh and Dukas, Helen, *Albert Einstein Creator & Rebel*, 1973, New York: A Plume Book, pp. 83-84.
[13] Seelig Carl, *Albert Einstein: A documentary biography*, Translated to English by Mervyn Savill 1956, London: Staples Press, p. 87; Seelig Carl, *Albert Einstein; eine dokumentarische Biographie*, 1954, Zürich: Europa Verlag, pp. 102-103.
[14] Laue to Einstein, June 2, 1906, *CPAE*, Vol. 5, Doc. 37.
[15] Planck to Einstein, July 6, 1907, *CPAE*, Vol. 5, Doc. 47.

3. But things may perhaps be different when it comes to the following question: Does the absolute vacuum (the free ether) possess any atomistic properties? Judging by your remark (*Ann*. 23, p. 372, 1907) that the electromagnetic state in a [finite] portion of space is determined by a *finite* number of quantities, you seem to answer this question in the affirmative, while I would answer it, at least in line with my present view, in the negative. For I do not seek the meaning of the quantum of action (light quantum) in the vacuum but at the sites of absorption and emission, and assume that the processes in the vacuum are described *exactly* by Maxwell's equations".

In spring 1906 Einstein wrote a review of Planck's lectures delivered at the University of Berlin during the winter semester of 1905-1906. Einstein published the review in *Beiblätter zu den Annalen der Physik*.[16] Einstein spoke in his review about "the chosen elements of area of finite size (= hν), where ν is the frequency and h is a universal constant, hν has the dimension of energy. The author repeats the need for the introduction of a universal constant h and emphasizes the importance of (not given in the book) the same physical interpretation".[17] This interpretation is not given in the book because it is Einstein's interpretation to which Planck objected.

Einstein did not introduce his "light quanta" heuristic into the relativity paper which he did not consider a revolutionary paper, but rather a natural development of classical electrodynamics and optics. Carl Seelig reported, "As opposed to several interpreters, Einstein would not agree that the relativity theory was a revolutionary event. He used to say: 'In the [special] relativity theory it is no question of a revolutionary act but of a natural development of lines which have been followed for centuries' ".[18]

In 1905 Planck was coeditor of the *Annalen der Physik* and soon a great supporter of Einstein's theory of relativity. Einstein used a seemingly conventional notion, "light complex", and he did not invoke his novel quanta of light heuristic with respect to the principle of relativity. He chose the language "light complex" for which no clear definition could be given in the relativity paper. But with hindsight, in 1905 Einstein made exactly the right choice not to mix concepts from his quantum paper with those from his relativity paper. He focused on the solution of his relativity problem, whose far-reaching perspectives Planck already sensed.

*I wish to thank Prof. John Stachel from the Center for Einstein Studies in Boston University for sitting with me for many hours discussing special relativity and its history.*

---

[16] *CPAE*, Vol. 2, Doc 37, and note 1.
[17] *CPAE*, Vol. 2, Doc 37, p. 766.
[18] Seelig, 1956, p. 82; Seelig, 1954, p. 97.